\documentclass[12pt]{iopart}
\usepackage{graphicx}
\usepackage{amsfonts}
\usepackage[english]{babel}
\usepackage{euscript}
\newcommand{\textfrac}[2]{{\textstyle\frac{#1}{#2}}}
\begin{document}

\title[Covering of the dKP--hyper CR Interpolating Equation]{%
A Covering for the dKP--hyper CR Interpolating Equation and Multi-Valued 
Einstein--Weyl Structures
}

\author{Oleg I. Morozov}

\address{Department of Mathematics, Moscow State Technical University of Civil Aviation, Kronshtadtskiy Blvd 20, Moscow 125993, Russia
\\
oim{\symbol{64}}foxcub.org}

\begin{abstract}
We apply the technique of integrable extensions to the symmetry pseudo-group of the dKP--hyper CR interpolating equation. This allows us to find a covering for this equation and to construct multi-valued Ein\-stein--Weyl structures.
%\keywords{Lie pseudo-groups \and Maurer--Cartan forms \and symmetries of differential equations 
%\and coverings of differential equations}
\end{abstract}


\maketitle

\section{Introduction}

We consider the equation
\begin{equation}
u_{yy}=u_{tx}+(b\,u_x-c\,u_y)\,u_{xx}+c\,u_x\,u_{xy},
\label{IE}
\end{equation}
$b=\mathrm{const}$, $c=\mathrm{const}$, found recently by M. Dunajski, \cite{Dunajski2008}, as a symmetry reduction of Pleba\~nski's second heavenly equation, \cite{Plebanski}. When  $c=0$ and $b\not = 0$, Eq. (\ref{IE}) co\-in\-ci\-des with the potential form of the Khokhlov--Zabolotskaya equation, \cite{KhokhlovZabolotskaya}, also known as dispersionless Kadomtsev--Petviashvili equation (dKP). For $b=0$ and $c\not = 0$ Eq. (\ref{IE}) becomes the hyperCR equation studied in \cite{Pavlov2003,Dunajski2004}. The Einstein-Weyl structure as\-so\-ci\-a\-ted to (\ref{IE}) is 
\begin{equation}
\left\{
\begin{array}{lcl}
h
&=&
(dy-c\,u_x\,dt)^2-4\,(dx+(c\,u_y-b\,u_x)\,dt)\,dt,
\\
\omega
&=&
-c\,u_{xx}\,dy+((c^2\,u_x+4\,b)\,u_{xx}- 2\,c\,u_{xy})\,dt. 
\end{array}
\right.
\label{EW_structure}
\end{equation}
A Lax pair
\begin{equation}
\left\{
\begin{array}{lcl}
q_t &=& \left((c\,z-b)\,u_x+c\,u_y+z^2\right)\,q_x+b\,(z\,u_{xx}+u_{xy})\,q_z,
\\
q_y &=& (c\,u_x+z)\,q_x+b\,u_{xx}\,q_z
\end{array}
\right.
\label{DLP}
\end{equation}
is proposed for Eq. (\ref{IE}) in \cite{Dunajski2008}. It contains %a 
differentiation with respect to a new in\-de\-pen\-dent variable $z$. The compatibility condition $q_{ty} = q_{yt}$ for system (\ref{DLP})  to\-ge\-ther with the requirement $u_{z}= 0$ coincides with Eq. (\ref{IE}). The equation $u_z = 0$ does not follow from (\ref{DLP}).

When $b\not =0$ and $c\not=0$, the simple scaling 
\begin{equation}
t=c^4\,b^{-3}\,\tilde{t}, 
\qquad 
x=c^2\,b^{-1}\,\tilde{x},
\qquad
y= -c^3\,b^{-2}\,\tilde{y}
\label{scaling}
\end{equation}
gives after dropping tildes the following equation:
\begin{equation}
u_{yy}=u_{tx}+(u_x+u_y)\,u_{xx}-u_x\,u_{xy}.
\label{main}
\end{equation}
The aim of this paper is to find a covering, \cite{KV84,KLV,KV89,KV99}, for Eq. (\ref{main}). We apply the method proposed in \cite{Morozov2008b} and find a contact integrable extension for the Cartan's struc\-ture equations of the symmetry pseudo-group of Eq. (\ref{main}). Integrating the extension yields a B\"acklund transformation and a covering equation for (\ref{main}). We use these results for constructing multi-valued Einstein--Weyl structures (\ref{EW_structure}) likewise in \cite{Morozov2008a} multi-valued solutions to dKP were found.

\section{Preliminaries}

\subsection{Coverings of PDEs}

Let $\pi_{\infty} : J^{\infty}(\pi) \rightarrow \mathbb{R}^n$ be the infinite jet bundle of local sections of the bundle $\pi : \mathbb{R}^n \times \mathbb{R} \rightarrow \mathbb{R}$. The coordinates on $J^{\infty}(\pi)$ are $(x^i, u_I)$, where
$I = (i_1,...,i_k)$ are symmetric multi-indices,  $i_1,...,i_k \in \{1,...,n\}$, $u_\emptyset= u$, and for any local section $f$ of $\pi$ there exists a section $j_{\infty}(f) : \mathbb{R}^n \rightarrow J^{\infty}(\pi)$ such that $u_I(j_{\infty}(f)) = \partial^{\# I}(f)/\partial x^{i_1} ... \partial x^{i_k}$, $\# I =\#(i_1,...,i_k) = k$.  The {\it total derivatives} on $J^{\infty}(\pi)$ are defined in the local coordintes as
\[
D_i = \frac{\partial}{\partial x^i}
+\sum \limits_{\# I \ge 0} u_{Ii} \, \frac{\partial}{\partial u_I}.
\]
\noindent
We have $[D_i, D_j] = 0$ for $i, j \in \{1,...,n\}$. A differential equation $F(x^i, u_K)=0$  defines a submanifold 
$\EuScript{E}^{\infty} = \{ D_I(F) =0 \,\,\vert\,\, \#I\ge 0\} \subset J^{\infty}(\pi)$,
where $D_I = D_{i_1}\circ ... \circ D_{i_k}$ for $I=(i_1,...,i_k)$. We denote restrictions of $D_i$ on $\EuScript{E}^{\infty}$ as $\bar{D}_i$.

In local coordinates, a {\it covering} over $\EuScript{E}^{\infty}$ is a bundle
$\widetilde{\EuScript{E}}^{\infty} = \EuScript{E}^{\infty} \times \EuScript{V} \rightarrow \EuScript{E}^{\infty}$ with fibre coordinates $v^\gamma$, $\gamma \in \{1,..., N\}$ or $\gamma \in \mathbb{N}$, equipped with {\it ex\-ten\-ded total derivatives}
\[
\widetilde{D}_i = \bar{D}_i
+\sum \limits_{\gamma}
T^{\gamma}_i (x^j, u_I, v^{\beta})\,\frac{\partial}{\partial v^\gamma}
\]
such that $[\widetilde{D}_i, \widetilde{D}_j ]=0$ whenever $(x^i, u_I) \in \EuScript{E}^{\infty}$.
Action of $\widetilde{D}_i$ on the fibre variables $v^\gamma$ gives  
\[
v_{x^i}^\gamma = T_i^\gamma (x^j, u_I, v^{\beta}).
\]
These equations are called a {\it B\"acklund transformation}. Excluding $u$ from them yields a system of {\sc pde}s for $v^\gamma$. This system is called a {\it covering equation}.

In terms of differential forms, the covering is defined by the forms, \cite{WE},
\[
\omega^{\gamma} = d v^{\gamma}- T^{\gamma}_i (x^j, u_I, v^{\beta})\,dx^i
\]
\noindent
such that
\begin{equation}
d \omega^{\gamma} \equiv 0 \,\,\,{\mathrm{mod}}\,\,\,\omega^{\beta}, \bar{\vartheta}_I 
\quad \Longleftrightarrow \quad (x^i,u_I) \in \EuScript{E}^{\infty},
\label{WE_def}
\end{equation}
where $\bar{\vartheta}_I$ are restrictions of contact forms 
$\vartheta_I = du_I-u_{I,k}\,dx^k$ on $\EuScript{E}^{\infty}$.

\subsection{Cartan's structure theory of Lie pseudo-groups}

Let $M$ be a manifold of dimension $n$. A {\it local diffeomorphism} on $M$ is a diffeomorphism 
$\Phi : \EuScript{U} \rightarrow \hat{\EuScript{U}}$ of two open subsets of $M$. A {\it pseudo-group} $\mathfrak{G}$ on $M$ is a collection of local dif\-feo\-mor\-phisms of $M$, which is closed under composition {\it when defined}, contains an identity and is closed under inverse. A {\it Lie pseudo-group} is a pseudo-group whose diffeomorphisms are local analytic solutions of an involutive system of partial differential equations. 

\'Elie Cartan's approach to Lie pseudo-groups is based on a possibility to characterize transformations from a pseudo-group in terms of a set of invariant differential 1-forms called {\it Maurer--Cartan} ({\sc mc}) {\it forms}. In a general case {\sc mc} forms $\omega^1$, ..., $\omega^m$ of a Lie pseudo-group $\mathfrak{G}$ on $M$ are defined on a direct product $\tilde{M} \times G$, where $\mu:\tilde{M}\times G \rightarrow M$ is a bundle, $m = \dim \,\tilde{M}$, $G$ is a finite-dimensional Lie group. The forms $\omega^i$ are semi-basic w.r.t. the natural projection $\tilde{M} \times G \rightarrow \tilde{M}$ and define a coframe on $\tilde{M}$, that is, a basis of the cotangent bundle of $\tilde{M}$. They characterize the pseudo-group $\mathfrak{G}$ in the following sense: a local diffeomorphism 
$\Phi : \EuScript{U} \rightarrow \hat{\EuScript{U}}$ on $M$ belongs to $\mathfrak{G}$ whenever there exists a local diffeomorphism 
$\Psi : \EuScript{V} \rightarrow \hat{\EuScript{V}}$ on $\tilde{M}\times G$ such that 
$\mu \circ \Psi = \Phi \circ \mu$ and the forms $\omega^j$ are invariant w.r.t. $\Psi$, that is,
\begin{equation}
\Psi^{*} \left(\omega^i\vert {}_{\hat{\EuScript{V}}} \right) 
= \omega^i\vert {}_{\EuScript{V}}.
\label{Phi_omega}
\end{equation}
Expressions of exterior differentials of the forms $\omega^i$ in terms of themselves give Cartan's structure equations of $\mathfrak{G}$:
\begin{equation}
d \omega^i = A_{\gamma j}^i \,\pi^\gamma \wedge \omega^j + B_{jk}^i\,\omega^j \wedge \omega^k,
\qquad B_{jk}^i = - B_{kj}^i.
\label{SE} 
\end{equation}
Here and below we assume summation on repeated indices. The forms $\pi^\gamma$, 
$\gamma \in \{1,...,\dim \, G\}$, are linear combinations of {\sc mc} forms of the Lie group $G$ and the forms $\omega^i$. The coefficients $A_{\gamma j}^i$ and $B_{jk}^i$ are either constants or functions of a set of invariants $U^\kappa : M \rightarrow \mathbb{R}$, 
$\kappa \in \{1,...,l\}$, $l < \dim\, M$, of the pseudo-group $\mathfrak{G}$, so 
$\Phi^{*} \left(U^{\kappa}\vert {}_{\hat{\EuScript{U}}} \right) 
= U^{\kappa}\vert {}_{\EuScript{U}}$
for every $\Phi \in \mathfrak{G}$. In the latter case, differentials of $U^\kappa$ are invariant 1-forms, so they are linear combinations of the forms $\omega^j$,
\begin{equation}
d U^\kappa = C_j^\kappa\,\omega^j,
\label{dUs}
\end{equation} 
where the coefficients $C_j^\kappa$ depend on the invariants $U^1$, ..., $U^l$ only.

Eqs. (\ref{SE}) must be compatible in the following sense: we have  
\begin{equation}
d(d \omega^i) = 0 = d \left(A_{\gamma j}^i \,\pi^\gamma \wedge \omega^j + B_{jk}^i\,\omega^j \wedge \omega^k \right),
\label{compatibility_conditions_SE} 
\end{equation}
therefore there must exist expressions 
\begin{equation}
d \pi^\gamma = W_{\lambda j}^\gamma\, \chi^\lambda \wedge \omega^j 
+ X_{\beta \epsilon}^\gamma\,\pi^\beta\wedge\pi^\epsilon
+Y_{\beta j}^\gamma\,\pi^\beta \wedge \omega^j
+Z_{jk}^\gamma\,\omega^j \wedge \omega^k,
\label{prolonged_SE}
\end{equation}
with some additional 1-forms $\chi^\lambda$ and the coefficients $W_{\lambda j}^\gamma$ to 
$Z_{j k}^\gamma$ depending on the invariants $U^\kappa$ such that the right-hand side of (\ref{compatibility_conditions_SE}) appear to be identically equal to zero after substituting for (\ref{SE}), (\ref{dUs}), and (\ref{prolonged_SE}). Also, from (\ref{dUs}) it follows that
the right-hand side of the equation
\begin{equation}
d(d U^\kappa) = 0 = d(C_j^\kappa\,\omega^j)
\label{compatibility_conditions_dUs}
\end{equation}
must be identically equal to zero after substituting for (\ref{SE}) and (\ref{dUs}).

The forms $\pi^\gamma$ are not invariant w.r.t. the pseudo-group $\mathfrak{G}$. Respectively, the structure equations (\ref{SE}) are not changing when replacing 
$\pi^\gamma \mapsto \pi^\gamma + z^\gamma_j\,\omega^j$ for certain parametric coefficients $z^\gamma_j$. The dimension $r^{(1)}$ of the linear space of these coefficients satisfies the fol\-lo\-wing inequality
\begin{equation}
r^{(1)} \le n\,\dim \, G  -  \sum \limits_{k=1}^{n-1} (n-k)\,\sigma_k, 
\label{CT}
\end{equation}
where the {\it reduced characters} $\sigma_k$ are defined by
\[
\sigma_k = \max \limits_{u_1,...,u_k}\, \mathrm{rank}\,\, \mathbb{A}_k(u_1,...,u_k)
\]
with the  matrices $\mathbb{A}_k$ inductively defined by
\[
\mathbb{A}_1(u_1) = \left(A^i_{\gamma j} \,u^j_1\right),
\qquad \mathbb{A}_l(u_1,...,u_l) = \left(
\begin{array}{c}
\mathbb{A}_{l-1}(u_1,...,u_{l-1})
\\
A^i_{\gamma j} \,u^j_l
\end{array}
\right),
\]
see \cite[\S 5]{Cartan1}, \cite[Def. 11.4]{Olver95} for the full discussion. 
The system of forms $\omega^k$ is {\it involutive}  if (\ref{CT}) is an equality, \cite[\S 6]{Cartan1}, \cite[Def. 11.7]{Olver95}.

Cartan's fundamental theorems, 
\cite[\S\S 16, 22--24]{Cartan1}, \cite{Cartan4}, 
\cite[\S\S 16, 19, 20, 25,26]{Vasil'eva},
\cite[\S\S 14.1--14.3]{Stormark}, 
state that for a Lie pseudo-group there exists a set of {\sc mc} forms whose structure equations satisfy the compatibility and involutivity conditions; conversely, if Eqs. (\ref{SE}), (\ref{dUs}) meet the compatibility conditions (\ref{compatibility_conditions_SE}), (\ref{compatibility_conditions_dUs}), and the involutivity condition, then there exists a collection of 1-forms $\omega^1$, ... , $\omega^m$ and functions $U^1$, ... , $U^l$ which satisfy (\ref{SE}) and (\ref{dUs}).  Eqs. (\ref{Phi_omega}) then 
define local diffeomorphisms from a Lie pseudo-group.

\vskip 5 pt
\noindent
{\bf Example 1.}
Consider the bundle $J^2(\pi)$ of jets of the second order of the bundle 
$\pi : \mathbb{R}^n \times \mathbb{R} \rightarrow \mathbb{R}^n$, 
$\pi : (x^1,...,x^n,u) \mapsto (x^1,...,x^n,u)$. A differential 1-form $\vartheta$ on $J^2(\pi)$ is called a {\it contact form} if it is annihilated by all 2-jets of local sections $f$ of the bundle $\pi$: $j_2(f)^{*}\vartheta = 0$. In the local coordinates every contact 1-form is a linear combination of the forms  $\vartheta_0 = du - u_{i}\,dx^i$, $\vartheta_i = du_i - u_{ij}\,dx^j$, $i, j \in \{1,...,n\}$, $u_{ji} = u_{ij}$. A local dif\-feo\-mor\-phism 
$\Delta : J^2(\pi) \rightarrow J^2(\pi)$, 
$\Delta : (x^i,u,u_i,u_{ij}) \mapsto (\hat{x}^i,\hat{u},\hat{u}_i,\hat{u}_{ij})$,
is called a {\it contact trans\-for\-ma\-tion} if for every contact 1-form $\hat{\vartheta}$ the form $\Delta^{*}\hat{\vartheta}$ is also contact. We denote by ${\rm{Cont}}(J^2(\pi))$  the pseudo-group  of contact transformations on $J^2(\pi)$.
Consider the following 1-forms 
\[
\Theta_0 = a\, \vartheta_0,
\quad
\Theta_i = g_i\,\Theta_0 + a\,B_i^k\,\vartheta_k,
\quad
\Xi^i =c^i\,\Theta_0+f^{ik}\,\Theta_k+b_k^i\,dx^k,
\]
\begin{equation}
\Theta_{ij} = a\,B^k_i\, B^l_j\,(du_{kl}-u_{klm}\,dx^m) + s_{ij}\,\Theta_0+w_{ij}^{k}\,\Theta_k,
\label{LCF}
\end{equation}
\noindent
defined on $J^2(\pi)\times\EuScript{H}$, where $\EuScript{H}$ is an open subset of $\mathbb{R}^{(2 n+1)(n+3)(n+1)/3}$ with local coordinates $(a$, $b^i_k$, $c^i$, $f^{ik}$, $g_i$, $s_{ij}$, $w^k_{ij}$, $u_{ijk})$, 
$i,j,k, \in \{1,...,n\}$,  such that $a\not =0$, $\det (b^i_k) \not = 0$, $f^{ik}=f^{ki}$, $s_{ij}=s_{ji}$, $w_{ij}^k=w_{ji}^k$,  $u_{ijk}=u_{ikj}=u_{jik}$, while $(B^i_k)$ is the inverse matrix for the matrix $(b^i_k)$.
As it is shown in \cite{Morozov2006}, the forms (\ref{LCF}) are {\sc mc} forms for ${\rm{Cont}}(J^2(\pi))$, that is, a local diffeomorphism
$\widetilde{\Delta} : J^2(\pi) \times \EuScript{H} \rightarrow J^2(\pi) \times \EuScript{H}$
satisfies the conditions
$\widetilde{\Delta}^{*}\, \hat{\Theta}_0 = \Theta_0$,
$\widetilde{\Delta}^{*}\, \hat{\Theta}_i = \Theta_i$,
$\widetilde{\Delta}^{*}\, \hat{\Xi}^i = \Xi^i$,
and $\widetilde{\Delta}^{*}\, \hat{\Theta}_{ij} = \Theta_{ij}$
if and only if it is projectable on $J^2(\pi)$, and its projection
$\Delta : J^2(\pi) \rightarrow J^2(\pi)$ is a contact transformation.
The structure equations for ${\rm{Cont}}(J^2(\pi))$ have the form
\begin{eqnarray}
\fl\hspace{10pt}
d \Theta_0 &=& \Phi^0_0 \wedge \Theta_0 + \Xi^i \wedge \Theta_i,
\nonumber
\\
\fl\hspace{10pt}
d \Theta_i &=& \Phi^0_i \wedge \Theta_0 + \Phi^k_i \wedge \Theta_k,
\nonumber
\\
\fl\hspace{10pt}
d \Xi^i &=& \Phi^0_0 \wedge \Xi^i -\Phi^i_k \wedge \Xi^k
+\Psi^{i0} \wedge \Theta_0
+\Psi^{ik} \wedge \Theta_k,
\nonumber
\\
\fl\hspace{10pt}
d \Theta_{ij} &=& \Phi^k_i \wedge \Theta_{kj} 
+ \Phi^k_j \wedge \Theta_{ki}
- \Phi^0_0 \wedge \Theta_{ij}
+ \Upsilon^0_{ij} \wedge \Theta_0
+ \Upsilon^k_{ij} \wedge \Theta_k + \Xi^k \wedge \Theta_{ijk},
\nonumber
\end{eqnarray}
where the additional forms $\Phi^0_0$, $\Phi^0_i$, $\Phi^k_i$, $\Psi^{i0}$, $\Psi^{ij}$,
$\Upsilon^0_{ij}$, $\Upsilon^k_{ij}$, and $\Theta_{ijk}$ depend on differentials of the coordinates of $\EuScript{H}$.

\vskip 5 pt
 
\noindent
{\bf Example 2.}
Suppose $\EuScript{E}$ is a second-order differential equation in one dependent and $n$ independent variables. We consider $\EuScript{E}$ as a submanifold in $J^2(\pi)$.
Let ${\rm{Cont}}(\EuScript{E})$ be the group of contact symmetries for $\EuScript{E}$. It consists of all the contact transformations on $J^2(\pi)$ mapping $\EuScript{E}$ to itself. Let 
$\iota_0 : \EuScript{E} \rightarrow J^2(\pi)$ be an embedding, and 
$\iota = \iota_0 \times \rm{id} : \EuScript{E}\times \EuScript{H} \rightarrow J^2(\pi)\times \EuScript{H}$. The {\sc mc} forms of ${\rm{Cont}}(\EuScript{E})$ can be computed from  the forms  $\theta_0 = \iota^{*} \Theta_0$,  $\theta_i= \iota^{*}\Theta_i$, $\xi^i = \iota^{*}\Xi^i$, and $\theta_{ij}=\iota^{*}\Theta_{ij}$ algorithmically by means of Cartan's method of equivalence, \cite{Cartan1,Cartan2,Cartan3,Cartan4,Gardner,Kamran,Olver95}, see details and examples in \cite{FelsOlver,Morozov2002,Morozov2006}.

\section{Cartan's structure of the symmetry pseudo-group for the dKP-hyperCR interpolating equation}

We use the method outlined in the previous section to compute {\sc mc} forms and structure equations of the pseudo-group of contact symmetries for Eq. (\ref{main}).  We write out here the structure equations for the forms $\theta_0$, $\theta_j$, $\xi^j$ with $j \in \{ 1,2,3 \}$ only; the full set of the structure equations is given in Appendix.  

We have
\begin{eqnarray}
%----------------------------------------------------------------------------------
\fl
d\theta_0
&=&
\eta_1 \wedge \theta_0
+\xi_1 \wedge \theta_1
+\xi_2 \wedge \theta_2
+\xi_1 \wedge \theta_1
+\xi_3 \wedge \theta_3,
\nonumber
\\
%----------------------------------------------------------------------------------
\fl
d\theta_1 
&=&
\textfrac{1}{2}\,\eta_1 \wedge \theta_1
+\textfrac{1}{4}\,\eta_2 \wedge (\theta_0+8\,\theta_3)
+\textfrac{1}{16}\,\left(
24\,(\theta_{22}-U\,\xi_1-\xi_2)-5\,\xi_3 \right) \wedge \theta_1
+2\,\theta_2 \wedge \theta_3
\nonumber
\\
\fl
&&
+\textfrac{1}{8}\,\left(
11\,\theta_2 
-16\,V\,\theta_{22}
+8\,\theta_{23}
+4\,(2\,V-1)\,\xi_2
+(8\,V-3\,U)\,\xi_3\right) \wedge \theta_0
+\xi_1 \wedge \theta_{11}
\nonumber
\\
\fl
&&
+\xi_2 \wedge \theta_{12}
+\xi_3 \wedge \theta_{13},
\nonumber
\\
%----------------------------------------------------------------------------------
\fl
d\theta_2
&=&
\textfrac{1}{16}\,\left(
8\,(\eta_1 
+\theta_{22}-U\,\xi_1-\xi_2)-3\,\xi_3 \right)\wedge \theta_2
+\xi_1 \wedge \theta_{12}
+\xi_2 \wedge \theta_{22}
+\xi_3 \wedge \theta_{23},
\nonumber
\\
%----------------------------------------------------------------------------------
\fl
d\theta_3
&=&
\eta_2 \wedge \theta_2
+\left(\eta_1 +\theta_{22}-U\,\xi_1-\xi_2
-\textfrac{1}{4}\,\xi_3\right) \wedge \theta_3
+\textfrac{5}{64}\,\left(8\,\theta_{22}-\xi_2+3\,\xi_3\right) \wedge \theta_0
+\xi_1 \wedge \theta_{13}
\nonumber
\\
\fl
&&
+\xi_2 \wedge \theta_{23}
+\xi_3 \wedge \theta_{12},
\nonumber
\\
%----------------------------------------------------------------------------------
\fl
d\xi_1
&=&
-\textfrac{1}{16}\,\left(
8\,\eta_1 +24\,(\,\theta_{22}+\xi_2)-5\,\xi_3
\right) \wedge \xi_1,
\nonumber
\\
%----------------------------------------------------------------------------------
\fl
d\xi_2
&=&
\textfrac{1}{8}\,\left(5\,\theta_0 
+8\,V\,\theta_2+8\,\theta_3-4\,U\,\xi_2\right) \wedge \xi_1
+\textfrac{1}{16}\,\left(8\,(\eta_1-\theta_{22})+3\,\xi_3\right) \wedge \xi_2
-\eta_2 \wedge \xi_3,
\nonumber
\\
%----------------------------------------------------------------------------------
\fl
d\xi_3
&=&
-(2\,\eta_2+\theta_2+U\,\xi_3) \wedge \xi_1
-(\theta_{22}-\xi_2) \wedge \xi_3, 
\nonumber
\end{eqnarray}
where the invariants
\begin{eqnarray}
\fl
U &=& u_{xx}^{-4}\,\left( (u_x+u_y)\,u_{xxx}^2+(u_{txx}-u_x\,u_{xxy}+u_{xx}\,(u_{xy}+4\,u_{xx})\,u_{xxx}
-u_{xxy}^2-u_{xx}^2\,u_{xxy}\right),
\nonumber
\\
\fl
V&=&u_{xxx}\,u_{xx}^{-2}
\nonumber
\end{eqnarray}
satisfy 
\begin{eqnarray}
%----------------------------------------------------------------------------------
\fl
dU
&=&
\textfrac{1}{2}\,U\,\eta_1+\eta_2+\eta_3
-\textfrac{5}{8}\,\theta_0
-\left(V+\textfrac{1}{8}\right)\,\theta_2
-\theta_3
-\left(4\,V-\textfrac{3}{2}\,U\right)\,\theta_{22}
-\theta_{23}
\nonumber
\\
\fl
&&
+\left(4\,V-U+\textfrac{1}{2}\right)\,\xi_2
-\left(V-\textfrac{11}{16}\,U\right)\,\xi_3,
\label{dU}
\\
%----------------------------------------------------------------------------------
\fl
dV&=&\textfrac{1}{2}\,V\,\left(
\eta_1
+\theta_{22}
+(6\,V-U)\,\xi_1
-\xi_2
\right)
-\textfrac{5}{16}\,V\,\xi_3,
\label{dV}
%----------------------------------------------------------------------------------
\end{eqnarray}
and where
\begin{eqnarray}
%----------------------------------------------------------------------------------
\fl
\theta_0 &=& V^2\,u_{xx}^3\,(du-u_t\,dt-u_x\,dx-u_y\,dy),
\nonumber
\\
%----------------------------------------------------------------------------------
%\fl
%\theta_1 &=& 
%V^2\,(V\,u_x+2\,u_{xxy}\,u_{xx}^{-2})\,du_y
%-V^2\,du_t 
%-V\,\left(u_{xxy}^2\,u_{xx}^{-4}+V^2\,(u_x+u_y)\right)\,du_x
%%
%\nonumber
%\\
%\fl
%&&
%%
%+
%V\,\left(
%(u_{tt}+u_x\,(u_{tx}-u_ty)+u_y\,u_{tx})\,V^2-2\,u_{ty}\,u_{xxy}\,V\,u{xx}^{-2}
%+u_{tx}\,u_{xxy}^2\,u_{xx}^{-4}
%\right)\,dt
%%
%\nonumber
%\\
%\fl
%&&
%%
%+V\,\left((u_{tx}-u_{x}\,u_{xy}+(u_x+u_y)\,u_{xx})\,V^2-2\,u_{xy}\,u_{xxy}\,V\,u_{xx}^{-2}
%+u_{xxy}^2\,u_{xx}^{-3}\right)\,dx
%%
%\nonumber
%\\
%\fl
%&&
%%
%-V\,\left(u_x\,(u_{tx}+(u_x+u_y)\,(u_{xx}-u_{xy})-u_{ty})\,V^2
%\right.
%%
%\nonumber
%\\
%\fl
%&&
%%
%\left.
%+2\,((u_x+u_y)\,u_{xx}^{-1}+(u_{tx}-u_x\,u_{xy})\,u_{xx}^2)\,u_{xxy}\,V
%-u_{xy}\,u_{xxy}\,q_{xx}^{-4}
%\right)\,dy
%%
%\nonumber
%\\
%\fl
%&&
%%
%+\left(V\,(2-u_{xy}\,u_{xx}^{-1})+u_{xxy}\,u_{xx}^{-2}\right)\,\theta_0
%\nonumber
%\\
%----------------------------------------------------------------------------------
\fl
\theta_2 &=& -V\,\left(du_x-u_{tx}\,dt - u_{xx}\,dx - u_{xy}\,dy\right),
\nonumber
\\
%----------------------------------------------------------------------------------
\fl
\theta_3 &=&
V\,u_{xxy}\,u_{xx}^{-2}\,du_x
-V^2\,du_y
+ V\,(V\,u_{ty}-u_{tx}\,u_{xxy}\,u_{xx}^{-2})\,dt
+V\,(V\,u_{xy}-u_{xxy}\,u_{xx}^{-1})\,dx
\nonumber
\\
\fl
&&
+V\,\left(V\,(u_{tx}+(u_x+u_y)\,u_{xx}-u_x\,u_{xy})-u_{xy}\,u_{xxy}\,u_{xx}^{-2}\right)\,dy
-\textfrac{5}{8}\,\theta_0,
\nonumber
\\
%----------------------------------------------------------------------------------
\fl
\theta_{22} 
&=& 
- u_{xx}^{-1}\,\left(du_{xx} - u_{txx}\,dt- u_{xxx}\,dx-u_{xxy}\,dy\right),
\nonumber
\\
%----------------------------------------------------------------------------------
\fl
\xi_1 
&=& 
u_{xx}\,V^{-1}\,dt, 
\label{MCFs}
\\
%----------------------------------------------------------------------------------
\fl
\xi_2 
&=&
u_{xx}^{-1}\,\left(u_{xxx}\,dx+u_{xxy}\,dy
-
(u_{xxx}\,(u_x+u_y)-u_{xxy}\,u_x-u_{xxy}^2\,u_{xxx}^{-1})\,dt
\right)
, 
\nonumber
\\
%----------------------------------------------------------------------------------
\fl
\xi_3 
&=& 
u_{xx}\,(u_x+2\,u_{xxy}\,u_{xxx}^{-1})\,dt+u_{xx}\,dy,
\nonumber
\\
%----------------------------------------------------------------------------------
\fl
\eta_1 
&=&
2\,u_{xxx}^{-1}\,du_{xxx}
+\left(10\,V-3\,U+4\,(V\,u_{xy}\,u_{xx}^{-1}-u_{xxy}\,u_{xx}^{-2})\right)\,\xi_1
+\textfrac{5}{8}\,\xi_3+3\,(\theta_{22}-\xi_2),
\nonumber
\\
%----------------------------------------------------------------------------------
\fl
\eta_2
&=&
-u_{xx}^{-2}\,du_{xxy}
-
\left(V^2\,(u_x+u_y+(u_{tx} -u_{xy}\,(u_x-1))\,u_{xx}^{-2})
\right.
\nonumber
\\
\fl
&&
\left.
+\left(4\,V\,u_{xxy}-\textfrac{3}{2}\,U\,u_{xxy}\right)\,u_{xx}^{-2}
-u_{xxy}^2\,u_{xx}^{-4}
\right)\,\xi_1
+\textfrac{1}{2}\,(3\,u_{xxy}\,u_{xx}^{-2}-1)\,\xi_2
\nonumber
\\
\fl
&&
-\left(V\,(u_{xy}\,u_{xx}+2)-\textfrac{11}{16}\,u_{xxy}\,u_{xx}^{-2}\right)\,\xi_3
-\textfrac{3}{2}\,u_{xxy}\,u_{xx}^{-2}\,\theta_{22}
+\textfrac{1}{2}\,u_{xxy}\,u_{xx}^{-2}\,\eta_1.
\nonumber
\end{eqnarray}

\section{Integrable extensions}

In \cite[\S 6]{BryantGriffiths}, the definition of integrable extension of an exterior differential system is designed to study finite-dimensional coverings. In general case, coverings of {\sc pde}s with three or more independent variables are infinite-dimensional, \cite{Marvan1992}. To cope with infinite-dimensional coverings we use a  natural generalization of the definition, \cite{Morozov2008b}. 

Suppose $\mathfrak{G}$ is a Lie pseudo-group on a manifold $M$ and $\omega^1$, ... , $\omega^m$ are its {\sc mc} forms with structure equations (\ref{SE}), (\ref{dUs}). Consider a system of equations
\begin{eqnarray}
\fl
d\tau^q &=& 
D^q_{\rho r} \, \eta^\rho \wedge \tau^r 
+ 
E^q_{r s} \, \tau^r \wedge \tau^s 
+
F^q_{r \beta} \, \tau^r \wedge \pi^\beta
+
G^q_{r j} \, \tau^r \wedge \omega^j
+
H^q_{\beta j} \, \pi^\beta \wedge \omega^j
\nonumber
\\
\fl
&+&
I^q_{j k} \, \omega^j \wedge \omega^k,
\label{extra_SE} 
\\
\fl
d V^\epsilon &=&  J^\epsilon_j \,\omega^j 
+  K^\epsilon_q \, \tau^q,
\label{dVs}
\end{eqnarray}
with unknown 1-forms $\tau^q$, $q \in \{1,...,Q\}$, $\eta^\rho$, $\rho \in \{1,...,R\}$,
and unknown functions $V^\epsilon$, $\epsilon \in \{1,...,S\}$ for some $Q, R, S \in \mathbb{N}$. The coefficients $D^q_{\rho r}$, ..., $K^\epsilon_q$ in (\ref{extra_SE}), (\ref{dVs}) are supposed to be functions of $U^q$ and $V^\gamma$.

\vskip 5 pt

\noindent
{\bf Definition 1.}
System (\ref{extra_SE}), (\ref{dVs}) is an {\it integrable extension} of system (\ref{SE}), (\ref{dUs}), if Eqs. (\ref{extra_SE}), (\ref{dVs}), (\ref{SE}), and (\ref{dUs}) together satisfy the compatibility and involutivity conditions. 

\vskip 5 pt

In this case from Cartan's third fundamental theorem for Lie pseudo-groups it follows that there exists a set of forms $\tau^q$ and functions $V^\epsilon$ which are solutions to Eqs. (\ref{extra_SE}) and (\ref{dVs}). Then $\tau^q$, $V^\epsilon$ together with $\omega^i$, $U^q$ define a Lie pseudo-group on a manifold $N \cong M \times \mathbb{R}^Q$.

\vskip 5 pt

\noindent
{\bf Definition 2.}
The integrable extension is called {\it trivial}, if there is a change of variables on $N$ such that in the new variables the coefficients $F^q_{r \beta}$, $G^q_{r j}$, $H^q_{\beta j}$, $I^q_{j k}$, and $J^\epsilon_j$ are equal to zero, while the coefficients $D^q_{\rho r}$, $E^q_{r s}$, and $K^\epsilon_q$ are independent of $U^q$. Otherwise, the integrable extension is called {\it non-trivial}.

\vskip 5pt 

Let $\theta^\gamma_I$ and $\xi^j$ be a set of {\sc mc} forms of the symmetry pseudo-group $\mathrm{Cont}(\EuScript{E})$ of a {\sc pde} $\EuScript{E}$ such that $\xi^1 \wedge ... \wedge \xi^n \not = 0$ on any solution manifold of $\EuScript{E}$, while $\theta_I^\gamma$ are contact forms. We take the following reformulation of the  definition (\ref{WE_def}) of a covering. 

\vskip 5 pt

\noindent
{\bf Definition 3.}  A non-trivial integrable extension of the form
\begin{equation}
d \omega^q =\Pi^q_r \wedge \omega^r + \xi^j \wedge \Omega^q_j 
\label{contact_ie}
\end{equation}
is called a {\it contact integrable extension} ({\sc cie}) of the structure equations of $\mathrm{Cont}(\EuScript{E})$ if

(1)\,\,\, $\Pi^q_r$ are some non-trivial differential 1-forms,

(2)\,\,\, $\Omega^q_j \equiv 0 \,\,\,{\mathrm{mod}}\,\,\,\theta^{\gamma}_I, \omega^q_j$ 
for some additional 1-forms $\omega^q_j$,

(3)\,\,\, $\Omega^q_j \not \equiv 0 \,\,\,{\mathrm{mod}}\,\,\, \omega^q_j$.

\vskip 7 pt

Since (\ref{contact_ie}) is integrable extension, Cartan's theorem yields existence of the forms $\omega^q$ satisfying (\ref{contact_ie}). From \cite[Ch. IV, Prop. 5.10]{BCGGG} it follows that the forms $\omega^q$ define a system of {\sc pde}s. This system is a covering for $\EuScript{E}$.

We apply this construction to the structure equations (\ref{main_SE}), (\ref{dU}), and (\ref{dV}) of the symmetry pseudo-group of Eq. (\ref{main}). We restrict our analysis to 
{\sc cie}s of the form 
\begin{eqnarray}
d \omega_0 
&=& 
\left(
\sum \limits_{i=0}^3 A_i \,\theta_i 
+ \sum {}^{*} B_{ij}\,\theta_{ij}
+ \sum \limits_{s=1}^7 C_s\,\eta_s 
+ \sum \limits_{j=1}^3 D_j\,\xi^j 
+ E\,\omega_1  
\right) \wedge \omega_0 
\nonumber
\\
&+& 
\sum \limits_{j=1}^3 \left(
\sum \limits_{k=0}^3 F_{jk}\,\theta_k + G_j\,\omega_1
\right) \wedge \xi^j,
\label{ie_main}
\end{eqnarray}
where $\sum {}^{*}$ means summation for all $i,j \in \mathbb{N}$ such that $1\le i \le j \le 3$, $(i,j)\not = (3,3)$, and consider two types of these {\sc cie}s:

\vskip 3pt
\noindent
TYPE 1 --- the coefficients $A_i$ to $G_j$ in (\ref{ie_main}) depend on the invariants $U$, $V$ of the symmetry pseudo-group of Eq. (\ref{main}) only;

\vskip 3pt
\noindent
TYPE 2 ---  the coefficients $A_i$ to $G_j$ depend also on one additional invariant, say $W$. In this case, the differential of this new invariant satisfies the following equation
\begin{equation}
d W = \sum \limits_{i=0}^3 H_i \,\theta_i 
+ \sum {}^{*} %\limits_{1\le i \le j \le 3,\,\, (i,j)\not = (3,3)} 
    I_{ij}\,\theta_{ij}
+ \sum \limits_{s=1}^7 J_s\,\eta_s 
+ \sum \limits_{j=1}^3 K_j\,\xi^j 
+ \sum \limits_{q=0}^1 L_q\,\omega_q,
\label{dW}
\end{equation}
where the coefficients $H_i$ to $L_q$ are functions of $U$, $V$, $W$.

\vskip 3 pt

The requirements of Defintions 1 and 3 yield over-determined systems for the coefficients $A_i$ to $G_j$ of the {\sc cie} of the first type  and $A_i$ to $L_q$ of the {\sc cie} of the second type. The results of analysis of these systems are summarized in the following theorem.

%=======================================================================================
\vskip 5 pt
\noindent
{\bf Theorem 1.} 
{\it 
There is no a {\sc cie} of the first type for the structure equations (\ref{main_SE}), (\ref{dU}), and (\ref{dV}). Every their {\sc cie} of the second type is contact-equivalent to the following one:}
\begin{eqnarray}
\fl
d\omega_0
&=&
\left(
\textfrac{1}{2} \,(\eta_1 - \theta_{22})
-\omega_1
-\textfrac{1}{2}(W^2+2\,(V-W)-U)\,\xi_1
-\textfrac{1}{16}\,(8\,W-11)\,\xi_3
\right)
\wedge
\omega_0
\nonumber
\\
\fl
&&
+\left(
W^2\,\omega_1
+\textfrac{5}{8}\,\theta_0
+W\,\theta_2
+\theta_3
\right)
\wedge
\xi_1
+\omega_1\wedge \xi_2
+\left(W\,\omega_1+\theta_2\right)\wedge\xi_3,
\label{extra_se}
\\
\fl
&&
~
\nonumber
\\
\fl
dW &=& 
V\,\omega_1
+\textfrac{1}{2}\,W\,\eta_1+\eta_2
+\textfrac{1}{2}\,W\,\theta_{22}
+\theta_2
+\textfrac{1}{2}\,W\,(V\,(W+4)-U)\,\xi_1
\nonumber
\\
\fl
&&
+\textfrac{1}{2}\,(V-W-1)\,\xi_2
+\textfrac{1}{16}\,(W\,(4\,V-5)+8\,V)\,\xi_3.
\label{d_W}
\end{eqnarray}
%=======================================================================================
\vskip 5 pt

Since the forms (\ref{MCFs}) are known, it is easy to find the form $\omega_0$ explicitly:

%=======================================================================================
\vskip 5 pt
\noindent
{\bf Theorem 2.} 
{\it
We have the following solution to Eq. (\ref{extra_se}) up to a contact equivalence:  
} 
\begin{eqnarray}
%%%%%%%%%%%%%%%%%%%%%%\fl
\omega_0
&=&
u_{xxx}\,u_{xx}^{-1}\,v_x^{-1}\left(
dv
-v_x\,\left(\ln^2 \vert v_x \vert -u_y-(\ln \vert v_x \vert +1)\,u_x+1\right)\,dt
\right.
\nonumber
\\
%
%%%%%%%%%%%%%%%%%%%%%%\fl
&&
\left.
-v_x\,dx
-v_x\,\left(\ln \vert v_x \vert -u_x\right)\,dy
\right).
\label{omega0}
\end{eqnarray}
\vskip 5 pt
%=======================================================================================

This form defines the following B\"acklund transformation or a Lax pair:
\begin{equation}
%\fl
\left\{
\begin{array}{lcl}
v_t &=& v_x\,\left(\ln^2 \vert v_x \vert -u_y-(\ln \vert v_x \vert +1)\,u_x+1\right),
\\
v_y &=& v_x\,\left(\ln \vert v_x \vert -u_x\right). 
\end{array}
\right.
\label{BT}
\end{equation}
It is easy to verify directly that Eqs. (\ref{BT}) are compatible whenever Eq. (\ref{main}) is satisfied. From Eqs. (\ref{BT}) it follows that 
\begin{equation}
%\fl
\left\{
\begin{array}{lcl}
u_x &=& \ln \vert v_x \vert - v_y\,v_x^{-1},
\\
u_y 
&=&
\left( v_y\,(\ln \vert v_x \vert +1)-v_t\right)\,v_x^{-1}-\ln \vert v_x \vert + 1,
\end{array}
\right.
\label{iBT}
\end{equation}
Cross-differentiating $u$ in this system yields  the covering equation:
\begin{equation}
v_{yy} =
v_{tx} +\left(
(v_y\,\ln \vert v_x \vert -v_t)\,v_x^{-1}+1
\right)\,v_{xx}
+ \left(
v_y\,v_x^{-1} -\ln \vert v_x \vert
\right)\,v_{xy}.
\label{CoveringEq}
\end{equation}

\section{Multi-valued Einstein--Weyl Structures}

We use the results of the previous section to construct a family of Einstein--Weyl structures (\ref{EW_structure}) depending on two arbitrary functions of one variable. We take the ansatz, \cite[Ch. VIII, \S\, 5.IV]{Bogoyavlenskiy}, \cite{Morozov2008a},
\begin{equation}
v_t = F(v_x), 
\qquad
v_y = G(v_x).
\label{ansatz}
\end{equation}
This system is compatible for every (smooth) functions $F$ and $G$. Substituting for (\ref{ansatz}) into (\ref{CoveringEq}) and denoting
\begin{equation}
v_x = s
\label{def_s}
\end{equation}
yields
\[
\fl
\qquad
\left(G^{\prime}(s)\right)^2 
=
F^{\prime}(s) 
+\left(
G(s)\,\ln \vert s \vert -F(s)
\right) \,s^{-1}
+1 
+\left(
G(s)\,s^{-1} - \ln \vert s \vert 
\right)\,G^{\prime}(s).
\]
We consider this as an {\sc ode} for the unknown fucntion $F$ and the functional parameter $G$ being an arbitrary smooth function.  Then we have
\begin{equation}
\fl
F(s) =
s\,\int s^{-1}\,\left(
\left(G^{\prime}(s)\right)^2+\ln \vert s \vert \,(G^{\prime}(s)-s^{-1}\,G(s))
-s^{-1}\,G(s)\,G^{\prime}(s)-1
\right)\,ds.
\label{integralF}
\end{equation}
From (\ref{ansatz}) and (\ref{def_s}) it follows that the function $s$ satisfies the  compatible system of {\sc pde}s 
\begin{equation}
s_t = F^{\prime}(s)\,s_x,
\qquad 
s_y = G^{\prime}(s)\,s_x.
\label{differentiated_ansatz}
\end{equation}
The general solution of this system in the implicit form reads
\begin{equation}
s = Q(x+t\,F^{\prime}(s)+y\,G^{\prime}(s)),
\label{s_multivalued}
\end{equation}
where $Q$  is an arbitrary (smooth) function of one variable. For $t=0$  and $y=0$ we have
$s=Q(x)$, so $Q$ is an initial value for Eqs. (\ref{differentiated_ansatz}). In general, Eq. (\ref{s_multivalued}) defines $s$ as a mul\-ti-valued function of $t$, $x$, and $y$.

Then Eqs. (\ref{iBT}), (\ref{ansatz}), and (\ref{def_s}) give 
\begin{equation}
\left\{
\begin{array}{lcl}
u_x
&=&
\ln \vert s \vert - s^{-1}\,G(s),
\\
u_y
&=&
\left(s^{-1}\,G(s)-1\right)\,\ln \vert s \vert - s^{-1}\,F(s)+1,
\end{array}
\right.
\label{EW_sol}
\end{equation}
where the function $F$ is defined by Eq. (\ref{integralF}), and the function $s$  is defined by
(\ref{s_multivalued}). Eqs. (\ref{EW_sol}) together with the scaling (\ref{scaling}) provide a family of Einstein-Weyl structures (\ref{EW_structure}) de\-pen\-ding on two arbitrary functions of  one variable.

Figures 1 to 4 show graphs of $u_x$ and $u_y$ at $t=-10$ and $t=10$ for the choice
of $G(s)=-3\,s$, $Q(x)=x^2+1$, and $F(s) = (5- \,\ln \vert s \vert )\,s$.

\begin{figure}
\centerline{\includegraphics*[angle=270,scale=0.5]{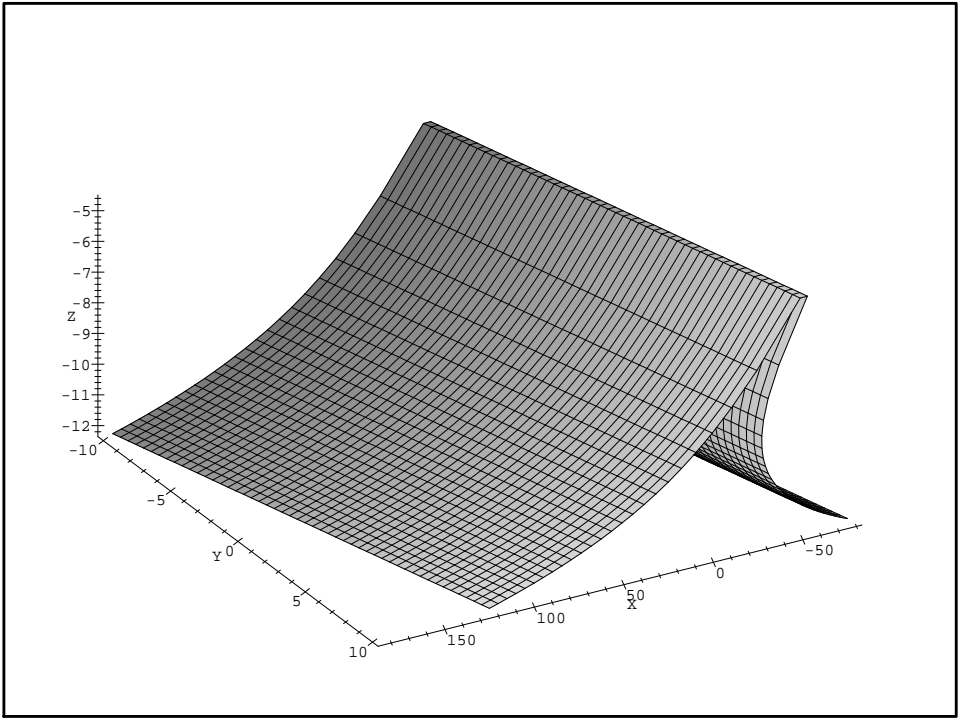}}
\caption{The graph of $u_x$ at $t=-10$.}
\label{plot1}
\end{figure}

\begin{figure}
\centerline{\includegraphics*[angle=270,scale=0.5]{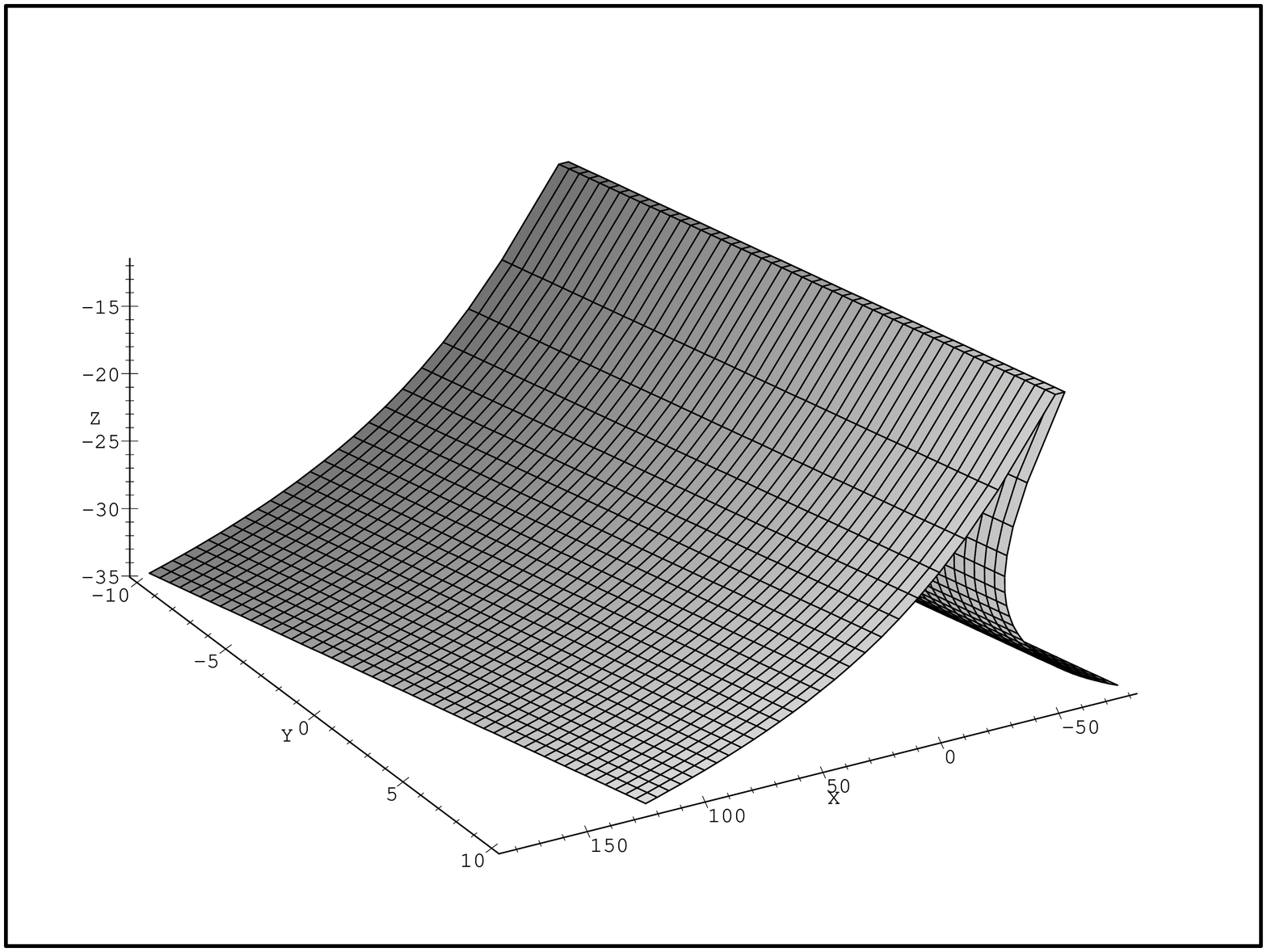}}
\caption{The graph of $u_y$ at $t=-10$.}
\label{plot2}
\end{figure}

\begin{figure}
\centerline{\includegraphics*[angle=270,scale=0.5]{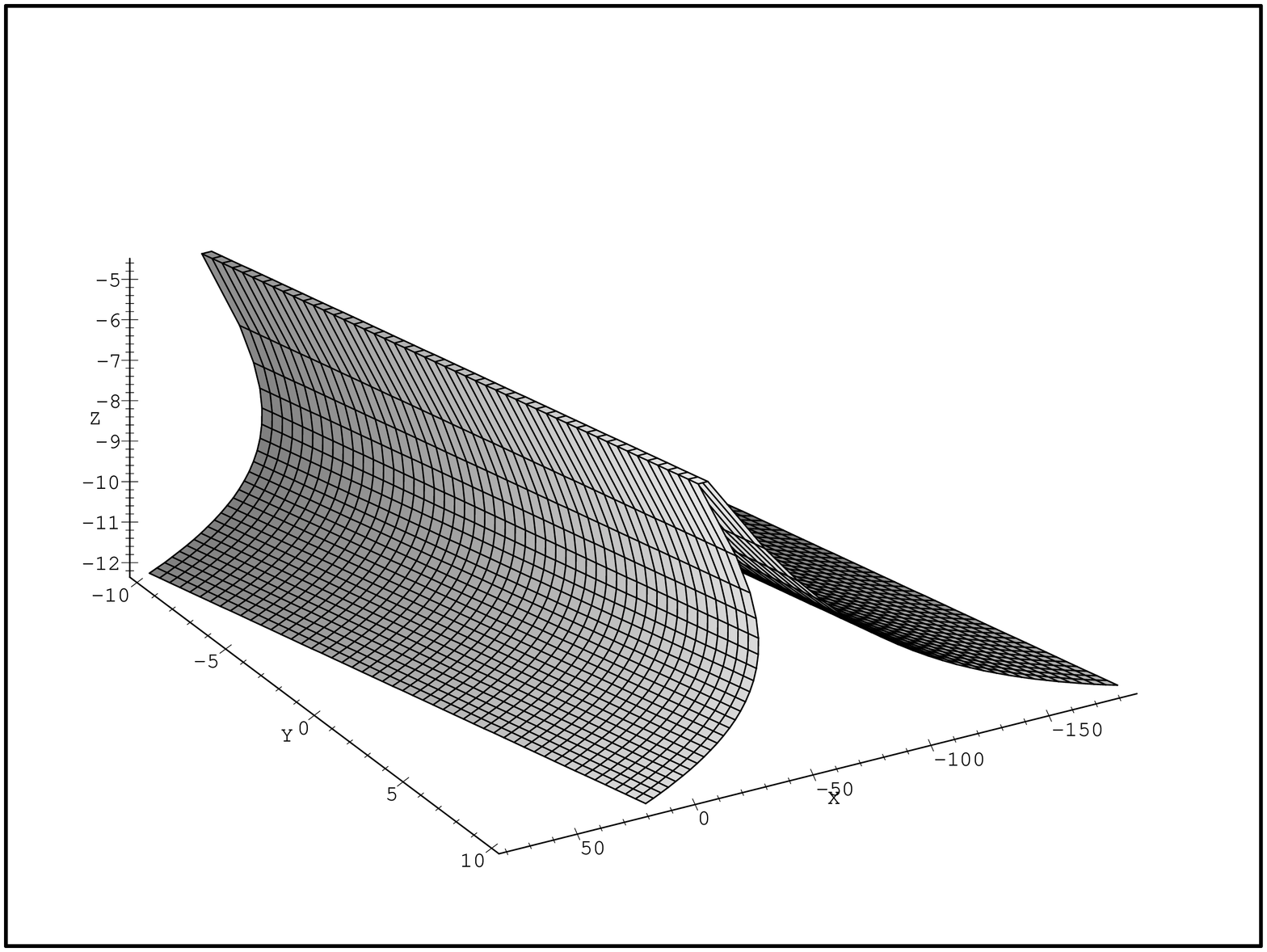}}
\caption{The graph of $u_x$ at $t=10$.}
\label{plot3}
\end{figure}

\begin{figure}
\centerline{\includegraphics*[angle=270,scale=0.5]{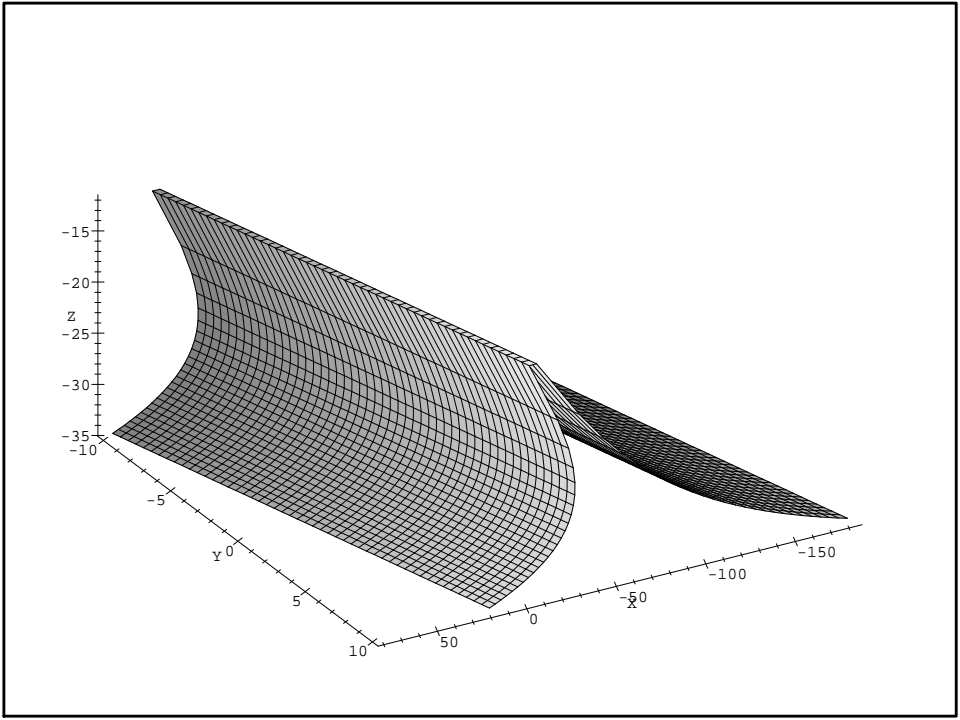}}
\caption{The graph of $u_y$ at $t=10$.}
\label{plot4}
\end{figure}

\section*{Acknowledgements}

I am grateful to the National Defense University of Taiwan, where the part of this work was done. My special thanks are to Jen-Hsu Chang for the warm hospitality in the NDU and valuable discussions, and to M.V. Pavlov for organizing my visit to Taiwan. The work was partially supported by the Russian-Taiwanese grant 95WFE0300007 (RFBR grant 06-01-89507-HHC) and the joint grant 09-01-92438-KE\_a of RFBR (Russia) and Consortium E.I.N.S.T.E.IN (Italy).

%%%%%%%%%%%%%%%%%%%%%%%%%%%%%%%%%%%%%%%%%%%%%%%%%%%%%%%%%%%%%%%%%%%%%%%%%%%%%%%%%%
\section*{Appendix}

The  structure equations of the symmetry pseudo-group for the dKP-hyperCR in\-ter\-po\-la\-ting equation read 

\begin{eqnarray}
%----------------------------------------------------------------------------------
\fl
d\theta_0
&=&
\eta_1 \wedge \theta_0
+\xi_1 \wedge \theta_1
+\xi_2 \wedge \theta_2
+\xi_1 \wedge \theta_1
+\xi_3 \wedge \theta_3,
\nonumber
\\
%----------------------------------------------------------------------------------
\fl
d\theta_1 
&=&
\textfrac{1}{2}\,\eta_1 \wedge \theta_1
+\textfrac{1}{4}\,\eta_2 \wedge (\theta_0+8\,\theta_3)
+\textfrac{1}{16}\,\left(
24\,(\theta_{22}-U\,\xi_1-\xi_2)-5\,\xi_3\right)  \wedge  \theta_1
+2\,\theta_2 \wedge \theta_3
\nonumber
\\
\fl
&&
+\textfrac{1}{8}\,\left(
11\,\theta_2 
-16\,V\,\theta_{22}
+8\,\theta_{23}
+4\,(2\,V-1)\,\xi_2
+(8\,V-3\,U)\,\xi_3\right) \wedge \theta_0
+\xi_1 \wedge \theta_{11}
\nonumber
\\
\fl
&&
+\xi_2 \wedge \theta_{12}
+\xi_3 \wedge \theta_{13},
\nonumber
\\
%----------------------------------------------------------------------------------
\fl
d\theta_2
&=&
\textfrac{1}{16}\,\left(
8\,(\eta_1 
+\theta_{22}-U\,\xi_1-\xi_2)-3\,\xi_3 \right)\wedge \theta_2
+\xi_1 \wedge \theta_{12}
+\xi_2 \wedge \theta_{22}
+\xi_3 \wedge \theta_{23},
\nonumber
\\
%----------------------------------------------------------------------------------
\fl
d\theta_3
&=&
\eta_2 \wedge \theta_2
+\left(\eta_1 +\theta_{22}-U\,\xi_1-\xi_2
-\textfrac{1}{4}\,\xi_3\right) \wedge \theta_3
+\textfrac{5}{64}\,\left(8\,\theta_{22}-\xi_2+3\,\xi_3\right) \wedge \theta_0
+\xi_1 \wedge \theta_{13}
\nonumber
\\
\fl
&&
+\xi_2 \wedge \theta_{23}
+\xi_3 \wedge \theta_{12},
\nonumber
\\
%----------------------------------------------------------------------------------
\fl
d\xi_1
&=&
-\textfrac{1}{16}\,\left(
8\,\eta_1 +24\,(\,\theta_{22}+\xi_2)-5\,\xi_3
\right) \wedge \xi_1,
\nonumber
\\
%----------------------------------------------------------------------------------
\fl
d\xi_2
&=&
\textfrac{1}{8}\,\left(5\,\theta_0 
+8\,V\,\theta_2+8\,\theta_3-4\,U\,\xi_2\right) \wedge \xi_1
+\textfrac{1}{16}\,\left(8\,(\eta_1-\theta_{22})+3\,\xi_3\right) \wedge \xi_2
-\eta_2 \wedge \xi_3,
\nonumber
\\
%----------------------------------------------------------------------------------
\fl
d\xi_3
&=&
-(2\,\eta_2+\theta_2+U\,\xi_3) \wedge \xi_1
-(\theta_{22}-\xi_2) \wedge \xi_3, 
\nonumber
\\
%----------------------------------------------------------------------------------
\fl
d\theta_{11}
&=&
\textfrac{3}{2}\,\eta_2 \wedge \theta_1
+\left(\left(2\,V -\textfrac{1}{2}\,U\right)\,\eta_2+2\,V\,\eta_3-\eta_4\right) \wedge \theta_0
+\eta_5 \wedge \xi_2
+\eta_6 \wedge \xi_3
+\eta_7 \wedge \xi_1 
\nonumber
\\
\fl
&&
+
\left(
\left(U-2\,V^2+\textfrac{51}{32}\,V\right)\theta_2
+(1-2\,V)\,\theta_3
+\textfrac{7}{4}\,\theta_{12}
-2\,V\,(U+3\,V)\,\theta_{22}
\right.
\nonumber
\\
\fl
&&
+\textfrac{1}{4}\,(6\,U-V)\,\theta_{23}
+\textfrac{1}{8}\,\left(U\,(24\,V-7)+V\,(48\,V-1)\right)\xi_2
\nonumber
\\
\fl
&&
\left.
+\textfrac{1}{16}\,\left(44\,V^2-15\,U^2+28\,U\,V\right)\xi_3
\right)
 \wedge \theta_0
+\left(2\,\theta_2
-11\,V\,\theta_{22}
+\theta_{23}
\right.
\nonumber
\\
\fl
&&
\left.
+\textfrac{1}{2}\,(22\,V-1)\,\xi_2
+\textfrac{1}{4}\,(4\,V+U)\,\xi_3\right) \wedge \theta_1
+(V\,\theta_{12}-3\,\theta_{13}) \wedge \theta_2
%=========================================================================
\label{main_SE}%===============================================================
%=========================================================================
\\
\fl
&&
+\left(\textfrac{3}{4}\,V\,\theta_2
+\theta_{12}
-2\,V\,\theta_{23}
+(U-2\,V)\,\xi_2 
-2\,V\,(U+V)\,\xi_3 
\right)
\wedge \theta_3
\nonumber
\\
\fl
&&
+\left(2\,\eta_1+3\,\theta_{22}-3\,\xi_2
-\textfrac{5}{8}\,\xi_3 \right)\wedge \theta_{11}
+2\,U\,\xi_2 \wedge \theta_{12}
+\left(4\,\eta_2+\textfrac{5}{2}\,U\,\xi_3\right) \wedge \theta_{13},
\nonumber
\\
%----------------------------------------------------------------------------------
\fl
d\theta_{12}
&=&
\left(\textfrac{1}{4}\,\eta_2 
-2\,V\,\theta_{22}
-\theta_{23}
+3\,V\,\xi_2 +\textfrac{1}{8}\,U\,\xi_3\right) \wedge \theta_2
+\eta_3 \wedge \xi_2
+\eta_4 \wedge \xi_3
+\eta_5 \wedge \xi_1
\nonumber
\\
\fl
&&
+\left(\eta_1+2\,\theta_{22}-2\,\xi_2-\textfrac{1}{2}\,\xi_3\right) \wedge \theta_{12}
-\left(\textfrac{5}{8}\,\theta_0 -\theta_3+U\,\xi_2\right) \wedge \theta_{22}
\nonumber
\\
\fl
&&
+\left(2\,\eta_2 +\textfrac{3}{2}\,U\,\xi_3\right) \wedge \theta_{23},
\nonumber
\\
%----------------------------------------------------------------------------------
\fl
d\theta_{13}
&=&
\textfrac{5}{64}\,\left(6\,\eta_2
-8\,(\eta_3-\theta_3-U\,\theta_{22}-\theta_{23})
+(8\,V-7)\, \theta_2
-6\,U\,(2\,\xi_2 +\xi_3)\right) \wedge \theta_0
+\eta_4 \wedge \xi_2
\nonumber
\\
\fl
&&
+\eta_5 \wedge \xi_3
+\eta_6 \wedge \xi_1
+\textfrac{5}{64}\,\left(
8\,(\theta_{22}-\xi_2)
+3\,\xi_3\right) \wedge \theta_1
+(3\,\eta_2+2\,\theta_2+2\,U\,\xi_3) \wedge \theta_{12}
\nonumber
\\
\fl
&&
-\left(\textfrac{7}{8}\,\theta_3
+\textfrac{1}{2}\,(2\,V-U)\,\xi_2
+V\,(U+V)\,\xi_3
\right) \wedge \theta_2
-\left(\theta_3 -\textfrac{3}{2}\,U\,\xi_2\right) \wedge \theta_{23}
\nonumber
\\
\fl
&&
+\textfrac{3}{8}\,\left(
2\,\eta_2
-4\,V\,(\theta_{22}-\xi_2)+U\,\xi_3 \right) \wedge \theta_3
+\textfrac{1}{16}\,\left(24\,\eta_1+40\,(\theta_{22}-\xi_2)-9\,\xi_3 \right)\wedge \theta_{13},
\nonumber
\\
%----------------------------------------------------------------------------------
\fl
d\theta_{22}
&=&
\eta_3 \wedge \xi_1
+\textfrac{1}{16}\,\left(
8\,(\eta_2-\theta_{22})+3\,\xi_3
\right)\wedge \xi_2
-\eta_2\wedge\xi_3,
\nonumber
\\
%----------------------------------------------------------------------------------
\fl
d\theta_{23}
&=&
\textfrac{1}{2}\,\eta_1 \wedge \theta_{23}
+\eta_2 \wedge (\theta_{22}-\xi_2)
+\eta_4 \wedge \xi_1
+\textfrac{1}{64}\,\left(
8\,(\theta_{22}-\xi_2)
+(64\,V-1)\,\xi_3
\right) \wedge \theta_2 
\nonumber
\\
\fl
&&
+\textfrac{1}{8}\,\left(8\,(\eta_3-\theta_3)
-5\,\theta_0
+12\,U\,\xi_2\right)\wedge \xi_3
+\textfrac{1}{8}\,\theta_{22}\wedge \left(
3\,(4\,\theta_{23}-\xi_2)-8\,U\,\xi_3\right)
\nonumber
\\
\fl
&&
+\textfrac{1}{16}\,\theta_{23} \wedge \left(24\,\xi_2-7\,\xi_3\right),
\nonumber
\\
%----------------------------------------------------------------------------------
\fl
d\eta_1
&=&
-\textfrac{1}{8}\,\left(
2\,\eta_2 
+6\,\theta_2-16\,(V\,\theta_{22}
+\theta_{23})
+4\,(2\,V-1)\,\xi_2
+(8\,V-3\,U)\,\xi_3
\right)
 \wedge \xi_1
\nonumber
\\
\fl
&&
-\textfrac{5}{8}\,(\theta_{22}-\xi_2) \wedge \xi_3,
\nonumber
\\
%----------------------------------------------------------------------------------
\fl
d\eta_2
&=&
\textfrac{1}{16}\,\left(
8\,(\eta_1+\theta_{22}-U\,\xi_1-\xi_2)
+3\,\xi_3
\right) \wedge \eta_2
+\textfrac{5}{16}\,\theta_0 \wedge \xi_1
-\textfrac{1}{8}\,\theta_2 \wedge 
\left(
3\,V\,\xi_1-\xi_3
\right)
\nonumber
\\
\fl
&&
+\textfrac{1}{2}\,\theta_3 \wedge \xi_1
+\left(
\theta_{12}+V\,\theta_{23}+V\,(V+U)\,\xi_3 
\right)\wedge \xi_1
+\textfrac{1}{2}\,\left(
\theta_{22}+(U-2\,V)\,\xi_1\right) \wedge \xi_2
\nonumber
\\
\fl
&&
+\left(
2\,V\,\theta_{22} +\theta_{23}
-\textfrac{1}{4}\,(8\,V+1)\,\xi_2
\right) \wedge \xi_3,
\nonumber
\\
%----------------------------------------------------------------------------------
\fl
d\eta_3
&=&
\textfrac{1}{16}\,\eta_1 \wedge \left(
8\,(\eta_3+V\,\theta_2+\theta_3)+5\,\theta_0-2\,U\,\xi_2
\right)
-\theta_{12}\wedge \left(
\left(V+\textfrac{9}{8}\right)\,\xi_1
+\xi_3
\right)
-\theta_{13} \wedge \xi_1
\nonumber
\\
\fl
&&
+\textfrac{1}{2}\,\eta_2 \wedge \left(
2\,\theta_2 +2\,(U-2\,V)\,\xi_1-\xi_2+U\,\xi_3
\right)
+\eta_4 \wedge \xi_1
-\textfrac{5}{8}\,\theta_1 \wedge \xi_1
\nonumber
\\
\fl
&&
+\textfrac{1}{16}\,\left(
24\,\theta_{22}+8\,(3\,U-8\,V)\,\xi_1-16\,\xi_2 -5\,\xi_3 \right)
\wedge \eta_3
+\textfrac{1}{32}\,\left(19\,U-8\,V\right)\,\xi_2 \wedge \xi_3
\nonumber
\\
\fl
&&
+\textfrac{5}{128}\,\theta_0 \wedge \left(
8\,\theta_{22}
+16\,(U-4\,V-1)\,\xi_1
-11\,\xi_3
\right)
-\textfrac{1}{2}\,\left(U^2-2\,U\,V-8\,V^2\right) \xi_1 \wedge \xi_3
\nonumber
\\
\fl
&&
+\textfrac{1}{16}\,\theta_2 \wedge \left(
8\,V\,\theta_{22}
+2\,((16\,V+9)\,U-(56\,V+9)\,V)\,\xi_1
-9\,\xi_2
+3\,V\,\xi_3
\right)
\nonumber
\\
\fl
&&
+\textfrac{1}{16}\,\theta_3 \wedge \left(
8\,\theta_{22}
+16\,(2\,U-4\,V-1)\,\xi_1
-11\,\xi_3
\right)
\nonumber
\\
\fl
&&
+\theta_{22}\wedge \left(
V\,(U-12\,V)\,\xi_1+\left(\textfrac{1}{4}\,U+V\right)\,\xi_2
\right)
\nonumber
\\
\fl
&&
+\textfrac{1}{2}\,\theta_{23}\wedge\left(
(U-2\,V)\,\xi_1-\xi_2-2\,V\,\xi_3
\right)
+\textfrac{1}{2}\,\left((U-2\,V)\,(U-1)-24\,V^2\right)\,\xi_1 \wedge \xi_2,
\nonumber
\\
%----------------------------------------------------------------------------------
\fl
d\eta_4
&=&
\eta_8 \wedge \xi_1
+\textfrac{1}{32}\,\left(
60\,\theta_0
+(96\,V-1)\,\theta_2 
+96\,\theta_3
+48\,U\,\theta_{22}
-24\,\theta_{23}
-96\,U\,\xi_2
\right.
\nonumber
\\
\fl
&&
\left.
+32\,(2\,V-U)\,\xi_3 
\right)
\wedge \eta_2
+\textfrac{1}{8}\,\left(
24\,\eta_2 
+7\,\theta_2
-12\,\theta_{23}
+3\,\xi_2
+4\,(7\,U-8\,V)\,\xi_3 \right)
\wedge \eta_3
\nonumber
\\
\fl
&&
+\left(
\eta_1 
+3\,(\theta_{22}-\xi_2)
-\textfrac{7}{4}\,\xi_3 
\right)
\wedge \eta_4
+\textfrac{1}{16}\,\left(
17\,U\,\theta_{22} +(24\,V-1)\,\theta_{23}-6\,(V+4\,U)\,\xi_2
\right) \wedge \theta_2
\nonumber
\\
\fl
&&
-\textfrac{5}{64}\,\left(
7\,(\theta_2-\theta_{22})+12\,\theta_{23}+4\,(\xi_2 -(7\,U-8\,V-2)\,\xi_3)
\right) \wedge \theta_0
\nonumber
\\
\fl
&&
+\textfrac{1}{8}\,\theta_3 \wedge \left(
7\,(\theta_2+\theta_{22})
-12\,\theta_{23}
-4\,\xi_2
-8\,(7\,U-8\,V-2)\,\xi_3
\right)
\nonumber
\\
\fl
&&
-\textfrac{1}{64}\,\left(
72\,(\theta_{22}+\xi_2)+73\,\xi_3
\right)\wedge \theta_{12}
+\textfrac{1}{128}\,\left(U\,(448\,V+143)-16\,V\,(32\,V^2+9)\right)\,\theta_2 \wedge \xi_3
\nonumber
\\
\fl
&&
+\left(
\left(V-\textfrac{11}{16}\,U\right)\,\xi_2-(U^2-13\,V^2)\,\xi_3
\right)  \wedge \theta_{22}
+\textfrac{1}{8}\,\left(
2\,(3\,U+1)\,\xi_2
+(U+4\,V)\,\xi_3
\right) \wedge \theta_{23}
\nonumber
\\
\fl
&&
+\textfrac{1}{4}\,(
4\,V\,(13\,V^2-1)-U\,(11\,U-8\,V-2)
)\,\xi_2 \wedge \xi_3,
\nonumber
\\
%----------------------------------------------------------------------------------
\fl
d\eta_5
&=&
\textfrac{3}{2}\,\eta_1 \wedge \eta_5
+\eta_2 \wedge \left(
4\,\eta_4+\textfrac{1}{2}\,\theta_{12}-2\,U\,\theta_{23}+(U-2\,V)\,\xi_2
\right)
+\left(
2\,\theta_2 -\xi_2+\textfrac{5}{2}\,U\,\xi_3
\right)
\wedge \eta_4
\nonumber
\\
\fl
&&
+\textfrac{1}{16}\,\left(
56\,(\theta_{22}-\xi_2)-13\,\xi_3 
\right)
\wedge \eta_5 
+\eta_8 \wedge \xi_3
+\eta_9 \wedge \xi_1
+\textfrac{5}{8}\,\theta_1 \wedge (\theta_{22}-\xi_2)
\nonumber
\\
\fl
&&
+\textfrac{5}{64}\,\left(
16\,\eta_3-(24\,V+1)\,\theta_2+16\,\theta_{12}
-8\,U\,(\theta_{22}+\theta_{23}+(U-4\,V+1)\,\xi_2)
\right)
\wedge \theta_0
\nonumber
\\
\fl
&&
+\textfrac{1}{16}\,\theta_2 \wedge \left(
16\,V\,\eta_3
-2\,(24\,V+1)\,\theta_3 
-8\,(4\,V+1)\,\theta_{12}
-48\,V\,(U-2\,V)\,\theta_{22}
\right.
\nonumber
\\
\fl
&&
\left.
-(U+V)\,\theta_{23}
+2\,\left(2\,U\,(2\,V+5)+V\,(16\,V-11)\right)\,\xi_2
+(U^2-4\,U\,V-4\,V^2)\,\xi_3 
\right)
\nonumber
\\
\fl
&&
+\left(2\,(\eta_3+\theta_{12})-\theta_{23}
+(4\,V-2\,U+1)\,\xi_2 
\right)
\wedge \theta_3
+\theta_{13} \wedge (\theta_{22}-\xi_2)
\nonumber
\\
\fl
&&
+\textfrac{1}{8}\,\left(
16\,(\eta_3+2\,V\,\theta_{22}-\theta_{23})
+(4\,U-16\,V+13)\,\xi_2 -4\,(4\,V-3\,U)\,\xi_3
\right)
\wedge \theta_{12}
\nonumber
\\
\fl
&&
+\left(\textfrac{1}{2}\,U^2+U\,V-12\,V^2\right)\,\theta_{22} \wedge \xi_2
+\textfrac{1}{2}\,
\left(
3\,(2\,V-U)\,\xi_2 
+\left(4\,V\,(U+V)-3\,U^2\right)\,\xi_3
\right)
\wedge \theta_{23}
\nonumber
\\
\fl
&&
-\textfrac{1}{2}\,\left(
2\,(3\,U-4\,V)\,\eta_3
+\left(8\,V^2+2\,U\,V-U^2\right)\,\xi_3
\right)
\wedge \xi_2,
\nonumber
\\
%----------------------------------------------------------------------------------
\fl
d\eta_6
&=&
\textfrac{1}{64}\,\eta_2 \wedge \left(
320\,\eta_5
+\left(80\,V-85\,U\right)\,\theta_0
-60\,\theta_1
+128\,V\,(U+V)\,\theta_2
+24\,U\,\theta_3
-160\,U\,\theta_{12} 
\right)
\nonumber
\\
\fl
&&
+\left(6\,V\,\theta_3 -\textfrac{5}{2}\,\theta_{13}\right)\wedge \eta_3
+\eta_4 \wedge \left(V\,\theta_2
+2\,(\theta_3-U\,\xi_2)\right)
-3\,\eta_5 \wedge (\theta_2 +U\,\xi_3)
\nonumber
\\
\fl
&&
+\left(
2\,\eta_1
+4\,\theta_{22}
-4\,\xi_2
-\textfrac{7}{8}\,\xi_3
\right) \wedge \eta_6
+\eta_8 \wedge \xi_2
+\eta_9 \wedge \xi_3
+\eta_{10} \wedge \xi_1
\nonumber
\\
\fl
&&
+\textfrac{5}{64}\,\left(
8\,(\eta_4+(5\,U-4\,V)\,\eta_3)
+10\,\theta_1
-2\,\theta_{12}
-4\,(5\,U+4\,V-2)\,\theta_3 
+2\,(7\,V-8\,U)\,\theta_{23}
\right.
\nonumber
\\
\fl
&&
\left.
-(96\,V^2+7\,U-10\,U^2+8\,U\,V-14\,V)\,\wedge \xi_2
+2\,(17\,V^2-5\,U^2+7\,U\,V)\,\xi_3
\right)
\wedge \theta_0 
\nonumber
\\
\fl
&&
+\textfrac{5}{128}\,\left(
32\,(\eta_3 -\theta_3) 
+8\,(U\,\theta_{22}-\theta_{23}+U\,\xi_2)
-9\,U\,\xi_3
\right)\wedge \theta_1
\nonumber
\\
\fl
&&
+\textfrac{1}{256}\left(
(U\,(400\,V+310)-V\,(640\,V+295))\,\theta_0
+40\,(8\,V+5)\,\theta_1 
\right)
\wedge \theta_2
\nonumber
\\
\fl
&&
+\textfrac{1}{32}\,\left(
(14\,U+192\,V^2+23\,V)\,\theta_2
-52\,\theta_{12}
+288\,V\,(2\,V-U)\,\theta_{22}
+8\,(3\,V-2\,U)\,\theta_{23}
\right.
\nonumber
\\
\fl
&&
\left.
+4\,(3\,U\,(16\,V-1)-6\,V\,(24\,V+1))\,\xi_2
+24\,V\, (V+U)\,\xi_3
\right) \wedge \theta_3
\nonumber
\\
\fl
&&
-
\textfrac{5}{64}\,\left(
8\,(\theta_{22}+\xi_2)-3\,\xi_3 
\right)\wedge\theta_{11}
+
\theta_{12}\wedge \left(\left(U+\textfrac{9}{8}\,V\right)\,\theta_2
-3\,V\,\theta_{23}
+2\,(U-2\,V)\,\xi_2
\right.
\nonumber
\\
\fl
&&
\left.
+2\,(U^2-2\,V^2-2\,U\,V)\,\xi_3
\right)
+\textfrac{1}{16}\,\theta_{13} \wedge \left(
25\,\theta_0 
+40\,(\theta_3+\theta_{23}) 
+(40\,V+23)\,\theta_2
\right.
\nonumber
\\
\fl
&&
\left.
-144\,V\,\theta_{22}
+4\,(36\,V-5\,U-3)\,\xi_2
+12\,(2\,V-U)\,\xi_3
\right)
+\textfrac{5}{8}\,V\,(U-12\,V)\,\theta_0 \wedge \theta_{22}
\nonumber
\\
\fl
&&
+\textfrac{3}{4}\,U\,(2\,V\,\theta_2-U\,\xi_2)\wedge \theta_{23}
+\textfrac{1}{2}\,V\,\theta_2 \wedge \left(
(U-6\,V)\,\xi_2-V\,(5\,U+12\,V)\,\xi_3
\right),
\nonumber
\\
%----------------------------------------------------------------------------------
\fl
d\eta_7
&=&
3\,\eta_2 \wedge \left(
2\,\eta_6-\textfrac{1}{4}\,\theta_{11}-\theta_{13}
\right)
-\eta_3\wedge \left(
13\,V\,\theta_1
-3\,\theta_{11}
\right)
+2\,\eta_4 \wedge (\theta_1-V\,\theta_3)
\nonumber
\\
\fl
&&
+\textfrac{1}{8}\,\eta_5 \wedge \left(
19\,\theta_0
+16\,(V\,\theta_2+\theta_3)
-20\,U\,\xi_2
\right)
-\textfrac{1}{2}\,\eta_6 \wedge (8\,\theta_2+7\,U\,\xi_3)
\nonumber
\\
\fl
&&
+\textfrac{1}{16}\,\left(
40\,\eta_1 
+72\,(\theta_{22}-\xi_2)
-15\,\xi_3
\right) \wedge \eta_7
-\eta_8 \wedge \theta_0
+\eta_9 \wedge \xi_2
+\eta_{10} \wedge \xi_3
+\eta_{11} \wedge \xi_1
\nonumber
\\
\fl
&&
+\textfrac{1}{32}\,\left(
4\,(13\,U^2-248\,V^2+4\,U\,V)\,\eta_2
-256\,V\,(U+V)\,\eta_3
+8\,(6\,U+7\,V)\,\eta_4
\right.
\nonumber
\\
\fl
&&
\left.
-5\,(52\,V+1)\,\theta_1
+(128\,V^3+V^2\,(4\,U-345)+56\,U^2+45\,U\,V)\,\theta_2
\right.
\nonumber
\\
\fl
&&
\left.
+8\,(16\,V\,(U+V)-3\,U)\,\theta_3
+60\,\theta_{11}
-(88\,U+59\,V)\,\theta_{12}
-72\,\theta_{13}
\right.
\nonumber
\\
\fl
&&
\left.
-32\,V\,(U^2-2\,U\,V-12\,V^2)\,\theta_{22}
+12\,V\,(U+2\,V)\,\theta_{23}
\right.
\nonumber
\\
\fl
&&
\left.
-2\,(192\,V^3+U^2\,(V-11)+8\,U\,V+4\,(U-3)\,V^2)\,\xi_2
\right.
\nonumber
\\
\fl
&&
\left.
-2\,(92\,U\,V^2-15\,U^3+6\,U^2\,V+360\,V^3)\,\xi_3
\right) \wedge \theta_0
+U^2\,\theta_{12} \wedge \xi_2
\nonumber
\\
\fl
&&
+\textfrac{1}{16}\,\left(
16\,(3\,U -4\,V)\,\eta_2
+4\,(52\,V-3)\,\theta_3 
-36\,\theta_{12}
-16\,V\,(20\,U-39\,V)\,\theta_{22}
\right.
\nonumber
\\
\fl
&&
\left.
+28\,V\,\theta_{23}
+2\,(3\,U\,(36\,V-1)-2\,V\,(156\,V-7))\,\xi_2
\right.
\nonumber
\\
\fl
&&
\left.
+\left(17\,(U^2-4\,V^2)+12\,U\,V\right)\,\xi_3 
\right)\wedge \theta_1
\nonumber
\\
\fl
&&
+\textfrac{1}{32}\,\left(
\left(117\,V-416\,V^2-88\,U\right)\theta_1 
+4\,(24\,V+23)\,\theta_{11}
-32\,V\,(U+3\,V)\,\theta_{12}
\right.
\nonumber
\\
\fl
&&
\left.
+8\,(4\,U+9\,V)\,\theta_{13}
\right)
\wedge \theta_2
%\right.
%
%\nonumber
%\\
%\fl
%&&
%
%\left.
+\textfrac{1}{4}\,\left(
16\,V\,(U+V)\,\eta_2
+9\,V\,(2\,U-V)\,\theta_2
+12\,\theta_{11}
\right.
\nonumber
\\
\fl
&&
\left.
-(4\,U+3\,V)\,\theta_{12}
-4\,\theta_{13}
-12\,V\,(U-2\,V)\,\theta_{23}
+2\,(U^2+4\,U\,V+12\,V^2)\,\xi_2
\right.
\nonumber
\\
\fl
&&
\left.
-4\,V\,(U^2-4\,U\,V-12\,V^2)\,\xi_3 
\right)
\wedge \theta_3
+\textfrac{3}{8}\,\left(
40\,V\,\theta_{22} 
-8\,\theta_{23} 
+4\,(U-10\,V+1)\,\xi_2
\right.
\nonumber
\\
\fl
&&
\left.
+(9\,U-8\,V)\,\xi_3
\right) \wedge \theta_{11}
+\left(
6\,V\,\theta_{23}
-3\,(U-2\,V)\,\xi_2
\right.
\nonumber
\\
\fl
&&
\left.
+\left(6\,V\,(U+V)-\textfrac{5}{2}\,U^2\right)\,\xi_3
\right)  \wedge \theta_{13}.
\nonumber
%----------------------------------------------------------------------------------
\end{eqnarray}

%%%%%%%%%%%%%%%%%%%%%%%%%%%%%%%%%%%%%%%%%%%%%%%%%%%%%%%%%%%%%%%%%%%%%%


\section*{Bibliography}
\begin{thebibliography}{99}
\bibitem{Bogoyavlenskiy} O.I. Bogoyavlenskiy,  Breaking So\-li\-tons. Nonlinear In\-te\-gra\-ble      Equa\-ti\-ons. Moscow, Nauka (1991)
\bibitem{BCGGG} Bryant, R.L., Chern, S.S., Gardner, R.B., Goldschmidt, H.L., Griffiths, P.A.:
    Exterior Differential Systems. N.Y., Springer-Verlag (1991)
\bibitem{BryantGriffiths} Bryant, R.L., Griffiths, Ph.A.:
    Characteristic cohomology of differential systems (II):
    conservation laws for a class of parabolic equations,
    Duke Math. J. {\bf 78}, 531--676  (1995)
\bibitem{Cartan1}  Cartan, \'E.: Sur la structure des groupes infinis de transformations.
    {\OE}uvres Compl{\`e}tes,   Part II, {\bf 2}, p. 571--715.
    Gauthier - Villars, Paris (1953)
\bibitem{Cartan2}  Cartan, \'E.: Les sous-groupes des groupes continus de transformations.
    {\OE}uvres Compl{\`e}tes,   Part II, {\bf 2}, p. 719--856.
    Gauthier - Villars, Paris (1953)
\bibitem{Cartan3} Cartan, \'E.: Les probl\`emes d'\'equivalence. 
    {\OE}uvres Compl{\`e}tes,   Part II, {\bf 2}, p. 1311--1334.
    Gauthier - Villars, Paris (1953)
\bibitem{Cartan4} Cartan, \'E.: La structure des groupes infinis. 
    {\OE}uvres Compl{\`e}tes,   Part II, {\bf 2}, p. 1335--1384.
    Gauthier - Villars, Paris (1953)
\bibitem{Dunajski2004} Dunajski, M.: A class of Einstein--Weil spaces associated to an integrable
   system of hydrodynamic type, J. Geom. Phys. {\bf 51}, 126-137 (2004)  
\bibitem{Dunajski2008} Dunajski, M.: An interpolating dispersionless integrable system. 
    J. Phys. A, Math. Theor., {\bf 41}, 315202 (2008)
\bibitem{Gardner} Gardner, R.B.: The method of equivalence and its applications.
    CBMS--NSF regional conference series in applied math., SIAM, Philadelphia (1989)
\bibitem{FelsOlver} Fels, M., Olver, P.J.: Moving coframes. I.
    A practical algorithm. Acta. Appl. Math. {\bf 51}, 161--213 (1998)
\bibitem{Kamran} Kamran, N.: Contributions to the Study of the Equivalence
    Pro\-blem of \'Elie Cartan and its Applications to Partial and
    Or\-di\-na\-ry Differential Equations.
    Mem. Cl. Sci. Acad. Roy. Belg., {\bf 45}, Fac. 7 (1989) 
\bibitem{KhokhlovZabolotskaya}  Zabolotskaya, E.A., Khokhlov, R.V.: Quasi-plane waves in the
     nonlinear acoustics of confined beams. Sov. Phys. Acoust., {\bf 15}, 35--40  (1969) 
\bibitem{KV84} Krasil'shchik, I.S., Vinogradov, A.M.: Nonlocal symmetries and the theory of
    coverings. Acta Appl. Math., {\bf 2}, 79--86  (1984)
\bibitem{KLV} Krasil'shchik, I.S., Lychagin, V.V., Vinogradov, A.M.:
    Geometry of Jet Spaces and Nonlinear Partial Differential Equations. Gordon and Breach, 
    New York (1986)  
\bibitem{KV89} Krasil'shchik, I.S., Vinogradov, A.M.: Nonlocal trends in the geometry of
    dif\-fe\-ren\-ti\-al equations: symmetries, con\-ser\-va\-ti\-on laws, and B\"acklund 
   trans\-for\-ma\-ti\-ons. Acta Appl. Math., {\bf 15}, 161--209  (1989)
\bibitem{KV99} Krasil'shchik, I.S., Vinogradov, A.M. (eds.):  Symmetries and
    Con\-ser\-va\-ti\-on Laws for Differential Equations of Ma\-the\-ma\-ti\-cal Physics.
    Transl. Math. Mo\-no\-graphs 182, Amer. Math. Soc., Pro\-vi\-dence (1999).
\bibitem{Marvan1992} Marvan, M.: On zero-curvature representations of partial differential
     equations.  Proc. Conf. on Diff. Geom. and Its Appl., Opava (Czech Republic),  103--122
     (1992)
\bibitem{Morozov2002} Morozov, O.I.: Moving coframes and symmetries of
    differential equations. J. Phys. A, Math. Gen., {\bf 35}, 2965--2977  (2002)
\bibitem{Morozov2006} Morozov, O.I.: Contact-equivalence problem for linear
    hyperbolic equations.  J. Math. Sci., {\bf 135}, 2680--2694  (2006)
\bibitem{Morozov2008a} Morozov, O.I.:  Cartan's structure theory of symmetry pseudo-groups,
   coverings and multi-valued solutions for the  Khokhlov--Zabolotskaya equation, 
   Acta Appl. Math. {\bf 101}, 231-241   (2008)
\bibitem{Morozov2008b} Morozov, O.I. Contact integrable extensions of symmetry pseudo-group 
   and coverings of the r-th modified dispersionless Kadomtsev--Petviashvili equation.
   {\tt arXiv:0809.1218v1 [math.DG]}  (2008)
\bibitem{Olver95} Olver, P.J.: Equivalence, Invariants, and Symmetry. Cambridge, Cambridge
    Uni\-ver\-si\-ty Press (1995)
\bibitem{Pavlov2003} Pavlov, M.V. Integrable hydrodynamic chains. J. Math. Phys., {\bf 44}, 
    4134--4156 (2003)
\bibitem{Plebanski} Pleba\~nski, J.F. Some solutions of complex Einstein equations. 
    J. Math. Phys., {\bf 16}, 2395--2402 (1975)
\bibitem{Stormark} Stormark, O.: Lie's Structural Approach to PDE Systems. Cambridge, Cambridge
    Uni\-ver\-si\-ty Press (2000)
\bibitem{Vasil'eva} Vasil'eva, M.V.: The Structure of Infinite Lie Groups of Transformations.
    Moscow, MGPI (1972) (in Russian)
\bibitem{WE} Wahlquist, H.D., Estabrook F.B.: Prolongation structures of nonlinear 
    evo\-lu\-ti\-on  equations.  J. Math. Phys., {\bf 16}, 1--7  (1975)
\end{thebibliography}
\end{document}